\documentclass[12pt]{iopart}
\usepackage{graphicx}
\usepackage[usenames]{color}
\usepackage{siunitx}
\begin{document}
\title[Reducing Poisson noise and baseline drift]{Reducing Poisson noise and baseline drift in X-ray spectral images with bootstrap Poisson regression and robust nonparametric regression}

\author{Feng Zhu$^1$, Binjie Qin$^*$$^1$, Weiyue Feng$^2$, Huajian Wang$^2$, Shaosen Huang$^1$, Yisong Lv$^3$, and Yong Chen$^4$}

\address{$^1$ School of Biomedical Engineering, Shanghai Jiao Tong University, Shanghai 200240, China}
\address{$^2$ CAS Key Laboratory of Nuclear Analytical Techniques and CAS Key Lab for Biomedical Effects of Nanomaterials and Nanosafety, Institute of High Energy Physics, Chinese Academy of Sciences, Beijing, 100049, China}
\address{$^3$ Department of Mathematics, Shanghai Jiao Tong University, Shanghai 200240, China}
\address{$^4$ School of Mechanical Engineering, Shanghai Jiao Tong University, Shanghai 200240, China}
\ead{$^*$ bjqin@sjtu.edu.cn}

\begin{abstract}
X-ray spectral imaging provides quantitative imaging of trace elements in biological sample with high sensitivity. We propose a novel algorithm to promote the signal-to-noise ratio (SNR) of X-ray spectral images that have low photon counts. Firstly, we estimate the image data area that belongs to the homogeneous parts through confidence interval testing. Then, we apply the Poisson regression through its maximum likelihood estimation on this area to estimate the true photon counts from the Poisson noise corrupted data. Unlike other denoising methods based on regression analysis, we use the bootstrap resampling method to ensure the accuracy of regression estimation. Finally, we use a robust local nonparametric regression method to estimate the baseline and subsequently subtract it from the X-ray spectral data to further improve the SNR of the data. Experiments on several real samples show that the proposed method performs better than some state-of-the-art approaches to ensure accuracy and precision for quantitative analysis of the different trace elements in a standard reference biological sample.
\end{abstract}


\maketitle

\section{Introduction}
\subsection{Various applications of X-ray spectral imaging}
X-ray spectral imaging has been used for more than half a century to identify and quantify the elemental composition of a wide variety of geological, biological, and medical target sample (Jenkins \etal 1995). Recently, due to the advent of the third generation synchrotron radiation facility, X-ray spectral imaging provides quantitative imaging of trace elements in biological sample with high sensitivity (sub-\si{\milli\gram\per\kilogram}) and high spatial resolution (sub-\si{\micro\meter} to \si{\nano\meter}). More and more researchers in the field of biomedicine and life science are showing great interest in this technology (Gherase and Fleming 2011). In the analysis of diseases such as Parkinson's disease and Alzheimer's disease, X-ray spectral images are useful when the quantitative imaging of element spatial distribution is needed to study the disease development (Popescu \etal 2009, Wang \etal 2010). Qin \etal (2011) used synchrotron radiation X-ray spectral images to explore the spatial association of copper in rat aortic media. Furthermore, as an interdisciplinary science complementary to genomics and proteomics, a new research subject called metallomics has been developed recently and is receiving great attention as a new frontier in the investigation of trace elements in biology (Mounicou \etal 2009). However, the accurate quantitative analysis is badly affected by the Poisson noise and baseline errors that are inherent in the X-ray spectral imaging. Especially, denoising X-ray spectral images that have low photon counts poses a big challenge in quantitative analysis of trace elements in biomedicine, which is also a focus of this study.

\subsection{Typical procedure of X-ray spectral imaging}
Different elements in a sample emit different scattered characteristic X-ray beams of many different energies when the sample is scanned and irradiated by the incident beams (such as X-ray, electrons) at every scanning location; each of these beams goes to the photon counting detector, and the intensities ({\it i.e.} photon counts at the detector) of these characteristic beams are proportional to the contents of the elements. This phenomenon is the very foundation of the analysis based on X-ray spectral imaging. Based on this physical law, a typical X-ray energy vs. the intensity spectrum divided by thousands of energy channels can be collected. Due to the physical nature of characteristic beams, only one or a few elements will be present at a particular energy channel of the spectrum when these elements are scanned at particular scanning location.

Scanning electron microscopy with an energy dispersive X-ray spectrometer (SEM-EDS) and energy dispersive micro synchrotron-based X-ray fluorescence (\si{\micro}SXRF) imaging are two commonly used methods to study the interactions of trace elements and single cells in natural system for the reason that they have relatively high sensitivity and high spatial resolution (Twining \etal 2003). Under energy-dispersive configuration, both methods use the same analytic procedure to quantitatively analyze the spectral images. The apparatus of SEM-EDS is cheaper and smaller than that of \si{\micro}SXRF imaging, but the monochromaticity, detection sensitivity and spatial resolution of \si{\micro}SXRF imaging are much higher than those of SEM-EDS (Van Grieken and Markowicz 2002). In this paper, the two data sets used in our experiments are produced by these two methods.

Before using these spectral imaging methods for the quantitative analysis in biomedicine, a specific analyte (or standard sample) in the form of either solid or solution is placed on the scanning platform to collect the multidimensional X-ray spectral data. The acquisition of multidimensional X-ray spectral data is a typical Poisson process (Boulanger \etal 2010), which makes the raw X-ray spectral data corrupted by Poisson noise. Besides, instrument-based systematic errors will also lead to a continuous, slowly varying baseline in the acquired spectral data (Twining \etal 2003). Fig. 1(a) displays such typical raw X-ray spectral data polluted with Poisson noise and systematic baseline. Many hardware-based efforts have been made to improve the signal-to-noise ratio (SNR) such as the insurance of $90^ \circ$ geometry between the incident and scattered beams (Geraki \etal 2004). In this paper, we proposed a novel software-based method that can reduce Poisson noise and baseline in the X-ray spectral data by means of signal processing. The desired effect of our method can be seen from pre-processing procedure in Fig. 1(b) with baseline (red line) subtracted from the denoised data (black line). The relatively pure spectral data (blue line) in Fig. 1(b) is the output of our proposed method.

The subsequent procedure is to calculate the intensities (photon counts) of characteristic X-ray beams of different elements from the spectrum. In this step, the signals of different elements need to be separated and then integrated (Fig. 1(c)). The original data (black dash line) has been separated as the sum of Ca (red peaks), Fe (green peaks), Cu (blue peaks) and Zn (purple peaks). Methods such as iterative least-squares fitting of a mathematical model combined with Monte Carlo simulations (Bekemans \etal 2003), baseline-corrected spectra fitting to a summed exponentially modified Gaussian (EMG) peak model with a sigmoidal baseline (Twining \etal 2003) are typically used in the separation of different elements' characteristic peaks. After the separation procedure, each element's peak areas are integrated to be equal to the number of photon counts for each element, which is then used for the quantitative mapping of element spatial distribution to disease development (Fig. 1(d)). The whole data processing procedure are displayed in Fig. 1.

Recently, a multi-platform open source software called PyMCA for the analysis of energy-dispersive X-ray fluorescence spectra has been developed (Sole \etal 2006). This software has combined many well performed algorithms in it, which can be used to calculate the intensities of characteristic X-ray beams (as displayed in Fig. 1(c)) in our standard biological samples. In next step (Fig. 1(d)), we demonstrate the use of these characteristic photon intensities so that we can obtain the quantitative amounts of different trace elements in standard biological samples (Wang \etal 2010).
\begin{figure}\centering
  \includegraphics[width=110mm]{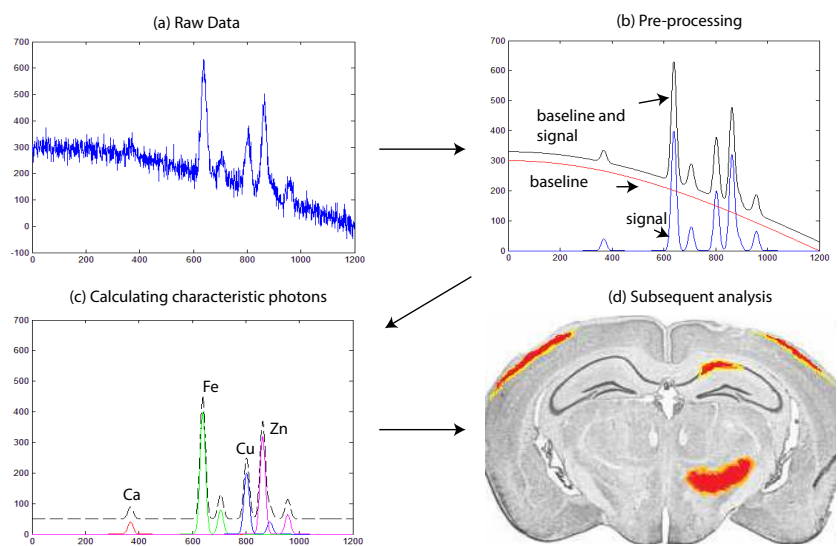}\\
  \caption{The typical processing procedure for X-ray spectral image data: (a) Simulated raw X-ray spectral data which are corrupted with Poisson noise and baseline;(b) Using preprocessing methods to reduce Poisson noise and baseline, baseline (red line) is subtracted from the denoised data (black line) and the clean signal (blue line) is obtained (here the baseline is lowered manually in order to give a clear show);(c) Separating different characteristic peaks and using these peaks to calculate the characteristic X-ray intensities of different elements, here we simulated the KL peak of Ca, KL and KM peaks of Fe, Cu, and Zn; the preprocessed data (black dash line) is raised manually in order to give a clear display;(d) the calculated X-ray intensities can be used in the subsequent analysis such as mapping certain elements in biological samples;}\label{1}
\end{figure}

\subsection{Review of Poisson denoising and baseline removal}
As has been mentioned above, we will describe a new signal processing method that deals with Poisson noise and baseline in the X-ray spectral image data. Our method is generally applicable to X-ray spectral images that have low photon counts and therefore pose a big challenge in quantitative analysis of trace elements in biomedicine.

There is an extensive literature on Poisson denoising methods which can be generally divided into three classes. The first use multiscale analysis technique (Zhang \etal 2008, Luisier and Blu 2008, Wang 2007, Spring and Clegg 2009) such as wavelet analysis. After the noisy signal being decomposed into noise and the useful signal by wavelet transform, the inversely transformed signals will be free from noise. Since the Poisson statistics are generally more difficult to be tackled than the Gaussian ones, the variance stabilizing transform is integrated into the multiscale analysis framework to transform the noise model from Poisson to Gaussian (Anscombe 1948, Spring and Clegg 2009, Makitalo and Foi 2011, Zhang \etal 2008, Palakkal and Prabhu 2012). In general, these algorithms usually require a prior knowledge of noise to set up appropriate parameters. The second class of methods estimate the true photon counts directly through statistical means, such as Bayesian inference combined with multiscale analysis (Timmerman and Nowak 1999, Kolaczyk 1999, Lefkimmiatis \etal 2009), hypothesis testing (Kolaczyk E D 2000), maximum likelihood estimation and regression analysis. In addition, there are variational approaches for Poisson denoising (Le \etal 2007, Chan and Chen 2007, Bonettini and Ruggiero 2011, Zhou and Li 2013). Regression analysis methods (Boulanger \etal 2008, Kervrann and Trubuil 2004) have a good performance when they deal with low photon count data. However, the statistical methods often give bad results when the sample size is small. Some resampling algorithms (Haynor and Woods 1989) have been introduced to overcome this small sample problem. Among these algorithms, bootstrap method (Dahlbom 2002) allows estimation of the sampling distribution of almost any statistic using only very simple methods. Considering the multidimensional X-ray spectral data are usually low photon count data containing a very high level of Poisson noise, we propose Poisson regression with bootstrap resampling methods to remove the Poisson noise in the multidimensional X-ray spectral data. To the best of our knowledge, we are the first to propose the bootstrap Poisson regression methods for X-ray spectral image denoising.

In order to further improve the performance of X-ray spectral imaging, we should also remove the baseline in the X-ray spectra. The baseline is caused by systematic errors such as the non-linear response of the detector (Van Grieken and Markowicz 2002). The continuous and low varying baseline in X-ray spectra can be treated as superposition on the original data (Ruckstuhl \etal 2001). Various methods in literature have been proposed to estimate the baseline of a spectrum. Liland \etal (2010) give an overview of the baseline correction for multivariate calibration of spectra. Here we choose the robust nonparametric regression methods based on the algorithm proposed by Ruckstuhl \etal (2001) for baseline removal since this method is both effective and robust. Being different from the original method, our algorithm uses a new regression kernel for the flexibility of second-order robust regression.

The remainder of this paper is organized as follows. Section 2 will explain the method in details. Section 2.2 introduces the bootstrap Poisson regression method. Section 2.3 outlines the robust regression methods for baseline removal. Section 3 gives the experimental results of our method in comparison with other state-of-the-art methods using standard samples including biological samples. In Section 4 we briefly summarize our method and future research directions.

\section{Methods}
\subsection{Data acquisition and data structure}
In this work, we use alloy wire data and two standard biological samples (bovine liver NIST 1577a and pig liver GBW 08551) as real samples in our multidimensional X-ray spectral imaging experiments. Both wire sample and biological samples are placed on an acquisition platform with micron spot of incident beam focused on the each scanning location in the samples. An energy dispersive X-ray spectrometer counts the characteristic photons emitted from each scanning sample point. Then the incident beam spot will move to the next scanning location for continuous data acquisition. The scanning time of each scanning point is the same. Based on the above description, the underlying structure of the acquired multidimensional X-ray spectral data can be taken as a 2D-1D structure. The 2D part of the multidimensional data refers to the two dimensional images that are formed from the total scanning points of the same energy channel while the 1D part refers to the whole spectrum for all energy channels at a single scanning point.

\subsection{Poisson denoising}
Photon counting X-ray spectral image data are typical Poisson distributed data. That is to say the photon count data $Y_i$ ($i=1,\cdots,N$, with $N$ being the total number of scanning points in each image of the 2D part of the data) follow a distribution with density function $f(Y_i;\lambda_i)=\frac{\lambda_i^{Y_i}\rme^{-\lambda_i}}{Y_i!}$, where $\lambda_i$ is the desired noise free photon counts, which can be estimated as mathematical expectation of $Y_i$. However, it is impossible to calculate the expectation of $f(Y_i;\lambda_i)$ ($\lambda_i$) through $Y_i$ itself. One method to solve the problem is to find more data from the same distribution ($f(Y_i;\lambda_i)$). As for X-ray spectral data, these identically distributed data can be found through the following analysis.

In modern high-resolution X-ray spectral imaging, an incident beam is scanned as a nanoscale spot in a raster pattern across the sample's surface to make the scanning points get fairly close to each other to offer sufficient details of the sample. The nanoscale spot is so small that the neighbouring scanning points around an interest point can be assumed to belong to a homogeneous region of the sample. In the biomedical imaging, the homogeneous regions are not equally extended to all directions from a point of interest, so that these irregular homogeneous regions usually introduce local discontinuities at some directions in the sample. Therefore, we need to check which neighbouring points belong to the same homogeneous part as the current scanning point of interest. A proper size of neighborhood should be determined so that there is sufficient number of homogeneous points that are chosen to recover the noise free photon counts $\lambda_i$. Furthermore, the bootstrap resampling method is used to produce sufficient candidates for good estimation of $\lambda_i$. With these spatial-domain local homogeneity assumption and sampling strategies, we have local photon count data that are independent and identically distributed (IID) for the following Poisson regression analysis.

By analyzing the above-mentioned spatial-domain features of the X-ray spectral data, we can use Poisson regression analysis to estimate the $\lambda_i$ from these IID data. Poisson regression assumes the response variable $Y_i$ has a Poisson distribution, and assumes the logarithm of its expected value ($\lambda_i$) can be modeled by a linear combination of unknown parameters. We suppose that the total number of homogeneous data in the local Poisson regression area is $m$, and these homogeneous data that belong to the same distribution as that of $Y_i$ in the neighbourhood of $Y_i$ are marked as $Y_{i,j}$ ($Y_i$ is included in $Y_{i,j},j=1,\cdots,m$). Therefore, the logarithmic forms of the expectation $\lambda_i$ for the $Y_{i,j}$ can be fitted with a linear model function:
\begin{equation}
    log(\lambda_i)=a_ix_j+b_i+\varepsilon_i; \lambda_i\approx\hat{\lambda_i}=\rme^{a_ix_j+b_i}, j=1,\cdots,m\label{eq1}
\end{equation}
where $x_j$ is an auxiliary explanatory variable for each point of $Y_{i,j}$, $\hat{\lambda_i}$ is the estimator of the true expectation $\lambda_i$, $a_i$ and $b_i$ are unknown parameters to be calculated, $\varepsilon_i$ is the white noise with a fixed variance and zero mean. The explanatory variable $x_j$ has no physics meaning but acts as a mathematic auxiliary tool for Poisson regression. To determine the explanatory variable $x_j$, the $Y_{i,j}$ are sorted increasingly or decreasingly so that the explanatory variable $x_j$ can simply be the sorting index. Therefore, the noise free photon counts $\lambda_i$ are approximated as:
\begin{equation}
    \lambda_i=E(Y_i)\approx\rme^{a_ix_{j'}+b_i}\label{eq2}
\end{equation}
where $x_{j'}$ is the corresponding explanatory variable to $Y_i$.

With the above-mentioned general scheme in mind, we should first choose the proper irregular homogeneous regions so that the $Y_i$'s neighbouring photon count data have the same distribution function as that of $Y_i$. Here we use the significance test to solve this problem. The acquired photon counts $Y_i$ are assumed to have a Poisson distribution. The data $Y_i^{'}$ in the $Y_i$'s neighbourhood should share the same cumulative distribution function:
\begin{equation}
    F(u)=\rme^{-Y_i}\sum_{k=0}^{u}{\frac{Y_i^k}{k!}}\label{eq3}
\end{equation}
With equation (3) we can calculate the lower and upper bound photon counts of the confidence interval of level 95\%. The data $Y_i^{'}$ that have the photon counts outside the confidence interval will be taken as the data samples having the different distributions from that of $Y_i$.

After choosing the homogeneous data that are candidate for choosing $Y_{i,j}$ in Poisson regression analysis of $Y_i$, the performance of estimator $\hat{\lambda_i}$ in equation (1) is then dependent on the $Y_{i,j}$'s size (or bandwidth) $m$ and can vary at each point of the photon count data sequence according to image contents. Small bandwidth will make local regression analysis sensitive to noises and outliers, while large bandwidth will create a large approximation error in local regression. In order to optimally estimate the bandwidth $m$, we analyze the performance of the estimator and consider the usual local $L_2$ risk (Boulanger \etal 2008) defined as:
\begin{equation}
    \mathcal{R}(\hat{\lambda_i},\lambda_i)=E[(\hat{\lambda_i}-\lambda_i)^2]\label{eq4}
\end{equation}
where $\lambda_i$ is the unknown expectation. The local risk $\mathcal{R}(\hat{\lambda_i},\lambda_i)$ is defined at each point and then differs from usual global performance measures that integrate errors on the whole images. This local risk of the candidates selected from the significance test reaches its minimum at each point. Choosing a new larger $m$ that does not increase the number of candidates means that the local discontinuity appears; the previous smaller $m$ is then considered as the optimal size. Otherwise, the local risk should be minimized to choose the optimal $m$ as follows.

Boulanger \etal (2008) have given in detail the solution of the minimization problem described in equation (4), which designs a sequence of increasing bandwidth: $M=\{m^{(n)}(x_j),n\in[0,N];m^{(n-1)}(x_j)<m^{(n)}(x_j)\}$. And then this sequence is used to detect the optimal bandwidth $m^{*}$ for the local smoothing:
\begin{equation}\label{eq5}
    {m^*}({x_j}) = \mathop {\sup }\limits_{{m^{(n)}}({x_j}) \in M} \{ n' < n:\left| {{{\hat \lambda_i }^{(n)}} - {{\hat \lambda_i }^{(n')}}} \right| < \vartheta {\upsilon _{n'}}(x)\}
\end{equation}
where $\vartheta=2\sqrt{2}$ is a positive constant, $\upsilon_{n^{'}}^2(x)$ is the variance of the data $Y_i$ in the regression context. Boulanger \etal (2008) have proved that the $m^{*}$ in the sequence of $M$ that satisfies equation (5) will be the optimal bandwidth $m$ that minimizes the risk described in equation (4). Typically, choosing eight or more homogeneous data points in the local regression area will ensure satisfactory estimation of $\lambda_i$.

In order to calculate the values of $a_i$ and $b_i$ in the equation (1), we use the principle of maximum likelihood estimation to compute the set of parameters $(a_i,b_i)$ that make the following log-likelihood function value as large as possible:
\begin{equation}\label{eq6}
    l(a_i,b_i)=\sum_{j=1}^{m}{Y_{i,j}(a_ix_j+b_i)-\rme^{a_ix_j+b_i}-log(Y_{i,j}!)}
\end{equation}
Unfortunately, directly computing $a_i$ and $b_i$ is difficult since equation (6) has no closed-form solution. An iterative weighted least squares method (Davison and Hinkley 1997) can be used to estimate $a_i$ and $b_i$. At each iteration an adjusted responses vector $\bi{z_i}=(\cdots,z_{i,j},\cdots)$ is regressed on the $x_j$ with elements $z_{i,j}$ being expressed as
\begin{equation}\label{eq7}
    z_{i,j} = w_j+(Y_{i,j}-\rme^{w_j})*\frac{1}{w_j}
\end{equation}
where the weight $w_j$ is given by the estimators $\hat a_i$ and $\hat b_i$ (corresponding to $a_i$ and $b_i$)
\begin{equation}\label{eq8}
    w_j=\hat{a}_ix_j+\hat{b}_i
\end{equation}
The results of each iteration given in the form of matrix is
\begin{equation}\label{eq9}
    \left[ {\begin{array}{*{20}{c}}
{{{\hat a}_i}}\\
{{{\hat b}_i}}
\end{array}} \right] = {({\mathbf{X}^T}\mathbf{W}\mathbf{X})^{ - 1}}{\mathbf{X}^T}\mathbf{X}{{\bi{z}}_i}
\end{equation}
where $\mathbf{X}$ is the matrix of $x_j$, $\mathbf{W}$ is the diagonal matrix of weights $w_j$ ({\it i.e.} $\mathbf{W}[s,t]={w_j}$, when $s=t$; $\mathbf{W}[s,t]=0$, when $ s\ne t$).

So far, the estimators $\hat a_i$ and $\hat b_i$ are ready to be substituted into the equation (2) for estimating the noise free photon counts $\lambda_i$. However, the accuracy of statistical estimation depends on the sample size. In practice, one needs as many samples as possible to ensure high estimation accuracy in statistical analysis. However, it is often not easy to obtain many samples. Efron (1979) has introduced bootstrap resampling methods to deal with this problem. Here we also use the bootstrap resampling methods to increase the accuracy of statistic analysis.

To apply bootstrap method to enhance the accuracy of estimating $\lambda_i$, we need to generate new sample data $Y_{i,j}^*$ from the original data $Y_{i,j}$. Specifically, we generate $Y_{i,j}^*$ from the estimated expectation ${\hat \lambda_i}$ by using following equation:
 \begin{equation}\label{eq10}
    Y_{i,j}^* = {\hat \lambda_i} + \hat \lambda_i^{1/2}*\varepsilon _j^*,j = 1, \cdots ,m
\end{equation}
where $\varepsilon _1^*, \cdots \varepsilon _m^*$ are adjustment parameters that determine the performance of the whole resampling method. It has been demonstrated that the residuals of Poisson regression can be used as these adjustment parameters for bootstrap resampling methods (Davison and Hinkley 1997). In order to deduce the residuals of Poisson regression, we first introduce the $\mathbf{H}$ matrix that is derived from equations (7)-(9):
\begin{equation}\label{eq11}
    \mathbf{H} = \mathbf{X}{({\mathbf{X}^T}\mathbf{W}\mathbf{X})^{ - 1}}{\mathbf{X}^T}\mathbf{W}
\end{equation}
With this definition of matrix $\mathbf{H}$, the standardized Pearson residuals can be written as
\begin{equation}\label{eq12}
    {r_{Pj}} = \frac{{{Y_{i,j}} - {{\hat \lambda_i}}}}{{{{\{ {{\hat \lambda_i }}(1 - {h_j})\} }^{1/2}}}},j = 1, \cdots ,m
\end{equation}
where $h_j$ is the $j$th diagonal element of the hat matrix $\mathbf{H}$.

The standardized Pearson residuals are further expressed as mean-adjusted Pearson residuals by ${r_{Pj}} - {\bar r_P}$, where $\bar r_P$ is the mean of $r_{Pj}$. These mean-adjusted Pearson residuals have all the qualities that bootstrap resampling method needs. With the adjustment parameters $\varepsilon _1^*, \cdots \varepsilon _m^*$ being sampled from these mean-adjusted, standardized Pearson residuals according to the bootstrap resampling rule, the new sample data $Y_{i,j}^*$ can be obtained as additional data to implement Poisson regression. Thus the desired expected value $\lambda_i$ (noise free photon counts in Equation (2)) can be estimated with higher accuracy. Theoretically the number of new samples generated from bootstrap methods needs to be infinity. In practice, 300 or more samples can ensure good results with pleasant accuracy.

\subsection{Baseline drift removal}
Baseline drift in X-ray spectral images is caused by hardware-based systematic errors such as the non-linear response of the detector. Based on the feature of slowly-varying local continuous baseline, it is convenient to remove baseline drift from the spectrum of each scanning point by using a robust local regression method (Ruckstuhl \etal 2001).

After the Poisson regression analysis, the spectral data of one scanning point can be modeled as follows:
\begin{equation}\label{eq13}
    V({c_k}) = g({c_k}) + s({c_k}) + {\varepsilon _k},k = 1, \cdots K
\end{equation}
where $V({c_k})$ is the processed data by Poisson regression, $g({c_k})$ is the baseline, $s({c_k})$ is the desired signal at the energy channel $c_k$, and ${\varepsilon _k}$ represents the measurement errors with zero mean and variance $\xi$, $K$ is the total number of points in the whole spectrum of a single scanning point.

In order to separate the three components in equation (13), a locally weighted scatter plot smoothing method (Cleveland 1979) can be used. With the data $V({c_k})$ at a energy channel $c_k$, we suppose the bandwidth of the local regression context around $c_k$ is $n$. In the $n$ defined local regression context, the Poisson regression processed data ${t_{k + i}}$ within this context can be defined with the following modified equation (from equation (13)):
\begin{equation}\label{eq14}
    {t_{k + i}} = g({c_{k + i}}) + {E_{k + i}};\quad {E_{k + i}} = s({c_{k + i}}) + {\varepsilon _{k + i}};\,i =  - n/2, \cdots ,n/2
\end{equation}
where $g({c_{k + i}})$ is the baseline function that has enough smoothness and $E_{k + i}$ is the sum of signal $s({c_{k + i}})$ and error ${\varepsilon _{k + i}}$. The baseline function $g(c)$ can be approximated by using second-order Taylor's formula at point $c_k$:
\begin{eqnarray}
    \eqalign{g(c) = g({c_k}) + g'({c_k})(c - {c_k}) + g''({c_k}){(c - {c_k})^2} + \Or ({(c - {c_k})^2})\cr
    g(c) \approx \hat g(c) = {\beta _0} + {\beta _1}(c - {c_k}) + {\beta _2}{(c - {c_k})^2}}\label{eq15}
\end{eqnarray}
where $\hat g(c)$ is the baseline estimated by the regression model. The parameter vector ${\bi{\beta }}({c_k}) = {[{\beta _0},{\beta _1},{\beta _2}]^T}$ can be calculated by incorporating a weight scheme into the local least-squares problem to decreases the influence of data points in proportion to their distance from ${c_k}$. That is,
\begin{eqnarray}
 \fl {\bi{\beta }}({c_k}) = \arg \mathop {\min }\limits_{\bi{\beta }} \sum\limits_{i =  - n/2}^{n/2} {K(\frac{{{c_{k + i}} - {c_k}}}{h})} \{ {t_{k + i}} - [{\beta _0} + {\beta _1}({c_{k + i}} - {c_k})\nonumber\\
     + {\beta _2}{({c_{k + i}} - {c_k})^2}]\}\label{eq16}
\end{eqnarray}
where $K[({c_{k + i}} - {c_k})/h]$ is a unimodal symmetric nonnegative weight function that is zero outside the ${c_k}$'s neighbourhood, which is defined by ${c_k} \pm h$ with $h$ being half of the bandwidth of the regression context. The estimation performance is not greatly dependent on the choice of the weight function $K$ (Ruckstuhl \etal 2001). Here we choose a logarithmic function to reduce the influence of the quadratic part in equation (15), so that the weight of data in the energy channels far from the current channel will drop quickly to have little effect on the estimation while the data in the close energy channels will have high weight:
\begin{equation}\label{eq17}
    K(u) = 1 - \log [(e - 1)u + 1]
\end{equation}
Equation (16) is used to estimate ${\bi{\beta }}({c_k})$ initially, next we use the residuals of this estimation to assign robustness weight ${w_r}({c_{k + i}})$ to each point, such that the points with large residuals will receive small robustness weights. The baseline curve $g(c)$ is then refined by performing a weighted least-squares fit, according to
\begin{eqnarray}
 \fl {\bi{\beta }}({c_k}) = \arg \mathop {\min }\limits_{\bi{\beta }} \sum\limits_{i =  - n/2}^{n/2} {{w_r}({c_{k + i}})K(\frac{{{c_{k + i}} - {c_k}}}{h})} \{ {t_{k + i}} - [{\beta _0} + {\beta _1}({c_{k + i}} - {c_k})\nonumber\\
     + {\beta _2}{({c_{k + i}} - {c_k})^2}]\}\label{eq18}
\end{eqnarray}
This fit is repeated iteratively to converge with the weights ${w_r}({c_{k + i}})$ always being calculated from the previous iteration. On the choice of various ${w_r}({c_{k + i}})$, we use Tukey's bisquare weights
\begin{equation}\label{eq19}
    {w_r}({c_{k + i}}) = \{ \max [1 - {({r_{k + i}}/b)^2},0]\}
\end{equation}
where ${r_{k + i}} = [{t_{k + i}} - \hat g({c_{k + i}})]/\sigma$. The $\sigma$ is estimated using the standardized median of absolute values of the residuals
\begin{equation}\label{eq20}
    \hat \sigma  = median(\left| {{t_{k + i}} - \hat g({c_{k + i}})} \right|)/0.6745
\end{equation}

The whole baseline estimation is summarized as follows:

(a) Equation (16) is used to compute $\hat g({c_k})$ initially;

(b) Use equation (19) to calculate the robustness weights ${w_r}({c_{k + i}})$;

(c) Use equation (18) to compute a new fitted value $\hat g({c_k})$.

The iterative steps (b) and (c) are repeated until $\left| {{{\hat g}_{n + 1}}({c_k}) - {{\hat g}_n}({c_k})} \right| < 0.01$, where ${\hat g_{n}}({c_k})$ and ${\hat g_{n + 1}}({c_k})$ are the results of step (c) at the $n$th and the $(n+1)$th iteration, respectively. Usually ten iterations will be enough.

The local regression bandwidth $n$ is the only parameter that needs to be set. For small size $n$, the robust local regression estimator is more likely to estimate $g({c_k}) + s({c_k})$ than $g({c_k})$. To avoid such a failure, a value $n$ being 2.5 times the full width of the widest peak in the spectra is recommended.

\section{Results and Analysis}
In this section, we firstly use alloy wire experiment data to test the performance of our method by comparing it with the four state-of-the-art algorithms. Then we perform quantitative analysis on two standard biological samples (bovine liver NIST 1577a and pig liver GBW 08551) by using our algorithm and the two best algorithms from the four state-of-the-art algorithms.
\subsection{Alloy wire experiment}
\subsubsection{Experimental data}

Alloy wire sample is described in figure 2 which consists of a series of six types of wires embedded in an epoxy matrix, with the wire alloys being composed from a pallet of six different elements. This alloy sample is imaged with SEM-EDS. Typical data acquisition conditions for this type of spectral image are described in detail by Kotula \etal (2003) and more details about this sample can be found in Keenan (2007). Figure 2(a) shows a standard SEM-EDS image of this sample together with the composition and component abundances. The image dimensions are $128 \times 128$ pixels, and a complete 1024-channel spectrum is collected at each pixel. Figure 2(b) shows a typical single-pixel spectrum for the Cu/Mn/Ni wire. The discrete nature of the data is clearly evident, and the SNR is sufficiently low so that the presence of Ni can not be clearly discriminated from the background.
\begin{figure}\centering
  \includegraphics[width=96.8mm]{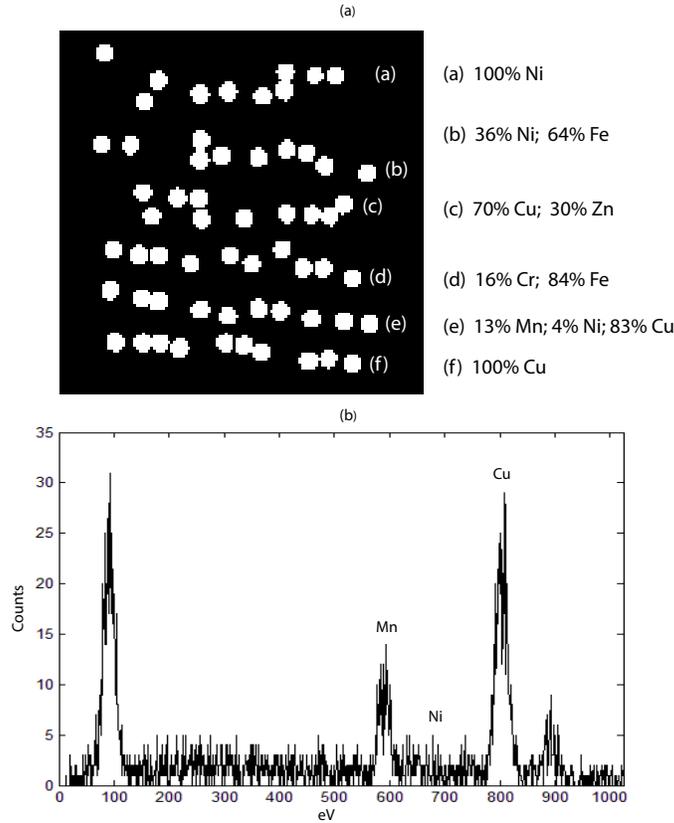}\\
  \caption{(a) The different positions of the wires with the different wire compositions and component concentrations: (a) 100\% Ni (b) 36\% Ni, 64\% Fe (c) 70\% Cu, 30\% Zn (d) 16\% Cr, 84\% Fe (e) 13\% Mn, 4\% Ni, 83\% Cu (f) 100\% Cu; ((a)-(f) are referred to the rows of dots in figure 2(a)). (b) A single-pixel spectrum from the Cu/Mn/Ni Wire.}\label{2}
\end{figure}

\subsubsection{Results and analysis}

In order to test the bootstrap Poisson regression with robust nonparametric regression baseline removal (BPR-RR) method, we introduce other four state-of-the-art methods for comparison purpose. The first method uses traditional Anscombe transform combined with Wiener filter (ATW) to remove the Poisson noise and local medians (LM) algorithm to remove the baseline artifacts (Friedrichs 1995). All parameters of this ATW-LM method are set as follows: the local filtering area of Wiener filter is $3\times3$; the window width of local medians is 200. The second method is designed for denoising mixed Poisson and Gaussian noise (MPG) (Zhang \etal 2008) and removing baseline by asymmetric least squares (ALS) methods (Eilers 2004). The parameters of MPG-ALS method are selected as: Poisson noise's weight alpha is set to 1; the mean and standard deviation of Gaussian component are 0 and 5; the number of iterations is set to 10. The SURE-LET (Luisier and Blu 2008) and BLS-GSM (Wang 2007) method are also introduced to denoise multidimensional X-ray spectral data. We set the parameters of these two methods as recommended in the original papers. The parameters of our proposed BPR-RR method are: The number of bootstrap samples is set to 300; the bandwidth of the local regression context for baseline estimation is set to 250. With those parameters all the methods mentioned above achieve their best performances.

\begin{figure}\centering
  \includegraphics[width=96.8mm]{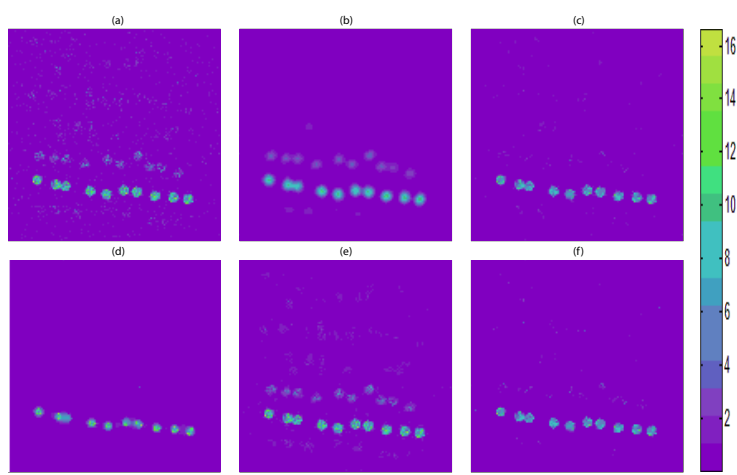}\\
  \caption{The X-ray spectral images at the energy channel of Mn's KL2 Line Energy (5.89 KeV) for (a) original data, (b) BLS-GSM method, (c) ATW-LM method, (d) MPG-ALS method, (e) SURE-LET method, (f) BPR-RR method.}\label{3}
\end{figure}
\begin{figure}\centering
  \includegraphics[width=96.8mm]{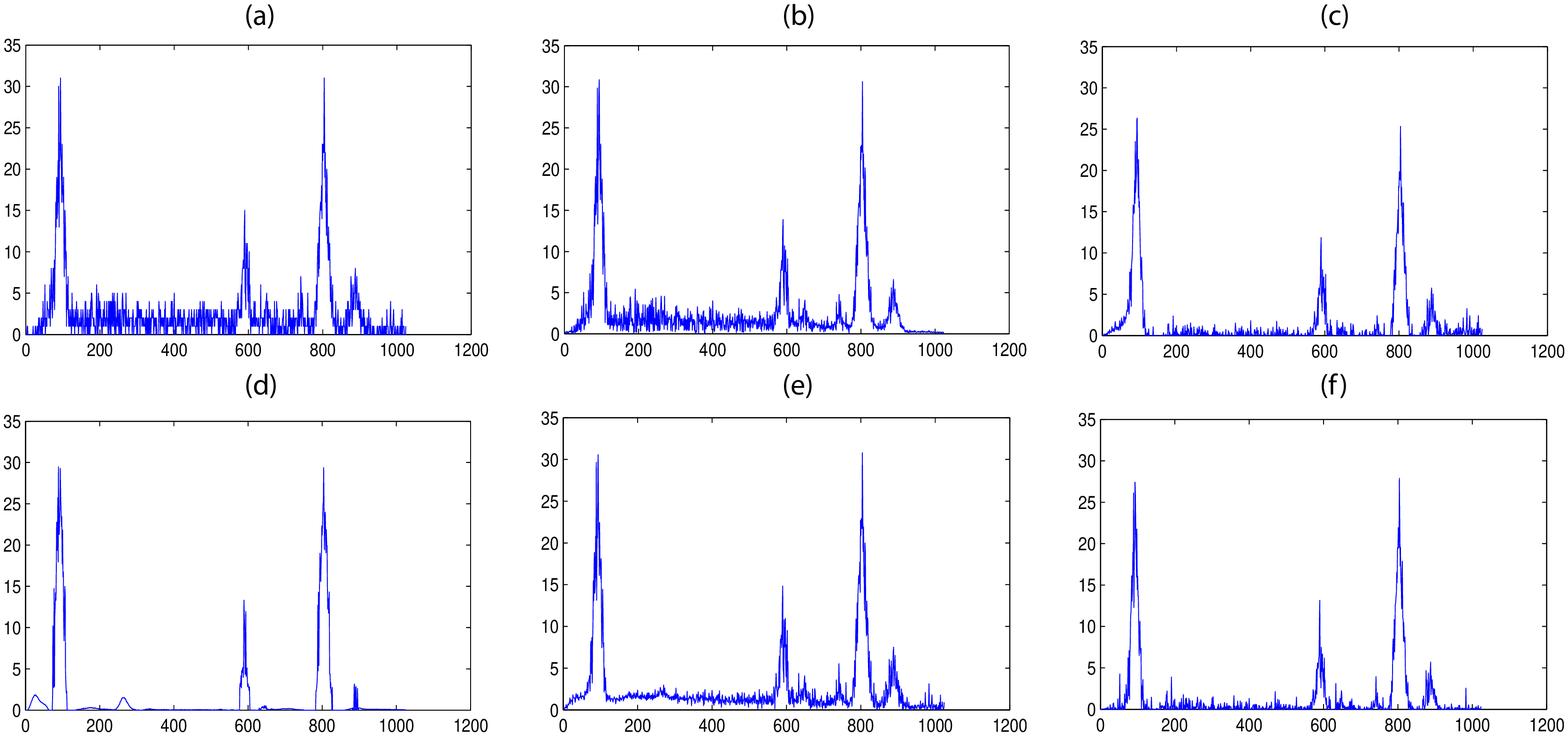}\\
  \caption{Typical single-pixel spectra of Mn/Ni/Cu alloy wire (line e in figure 2(a), the line displayed in figure 3) for (a) original data, (b) BLS-GSM method, (c) ATW-LM method, (d) MPG-ALS method, (e) SURE-LET method, (f) BPR-RR method (vertical axis: photon counts; horizontal axis: eV).}\label{4}
\end{figure}
\begin{figure}\centering
  \includegraphics[width=96.8mm]{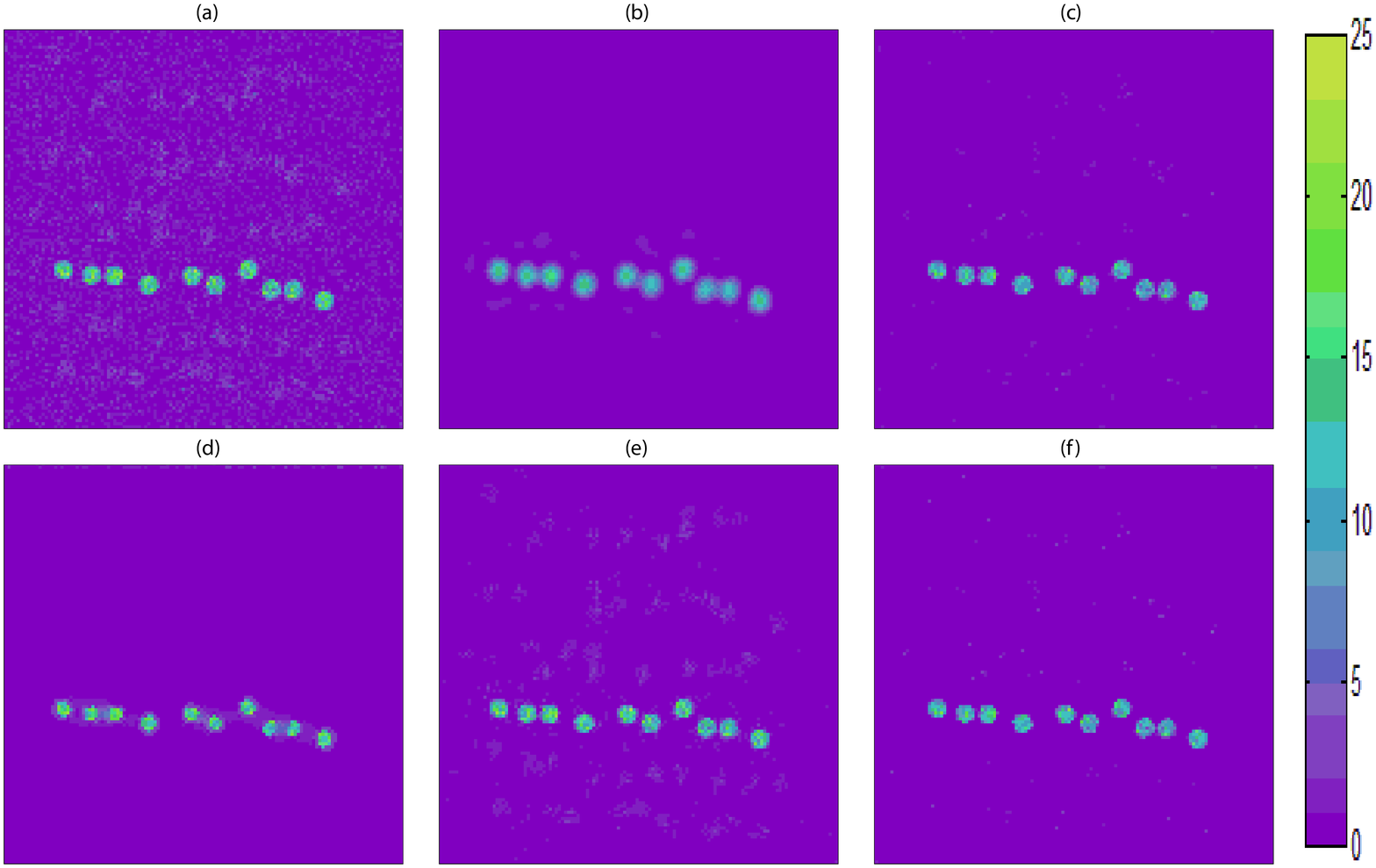}\\
  \caption{The X-ray spectral images at the energy channel of Cr's KL2 Line Energy (5.41 KeV) for (a) original data, (b) BLS-GSM method, (c) ATW-LM method, (d) MPG-ALS method, (e) SURE-LET method, (f) BPR-RR method.}\label{5}
\end{figure}
\begin{figure}\centering
  \includegraphics[width=96.8mm]{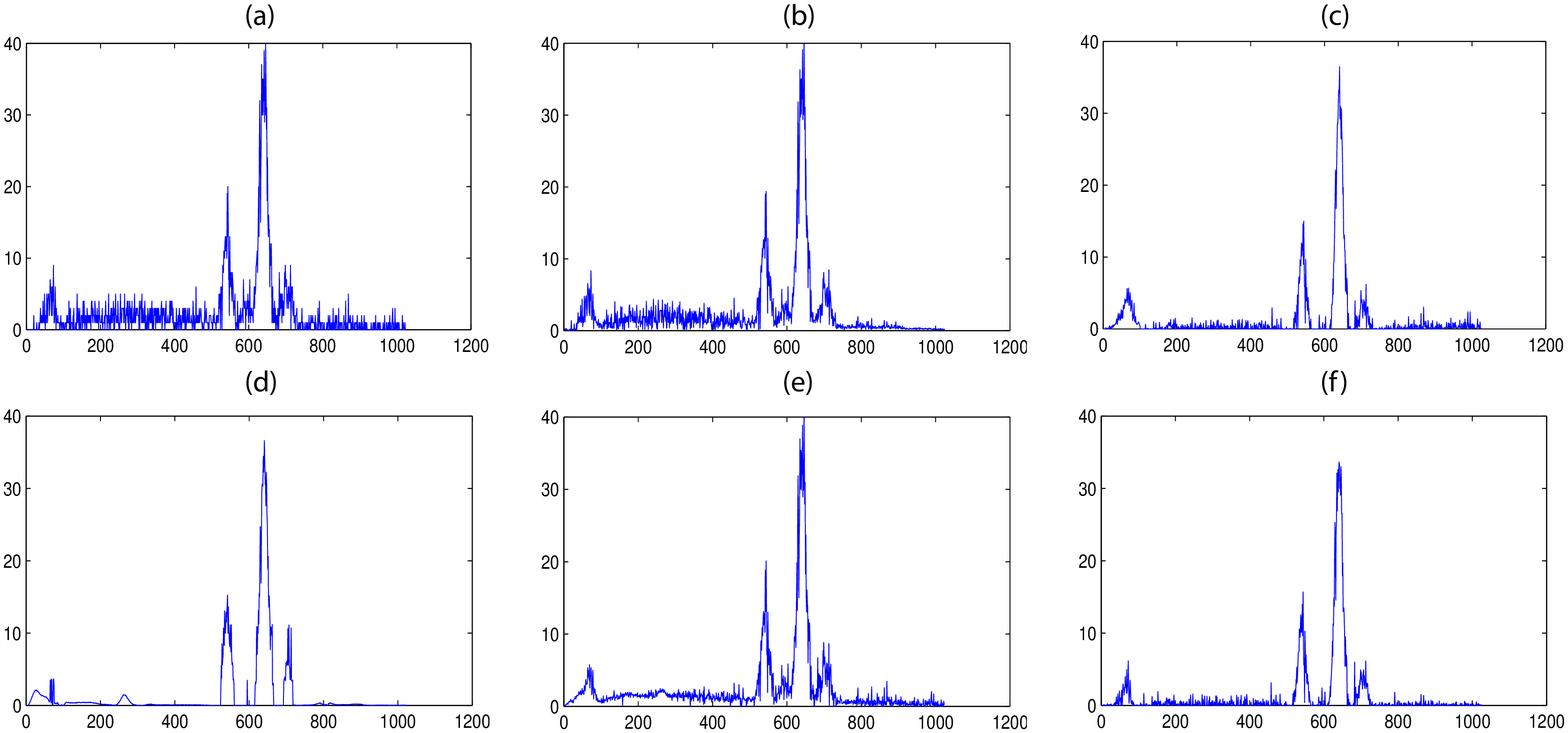}\\
  \caption{Typical single-pixel spectra of Cr/Fe alloy wire (line d in figure 2(a), the line displayed in figure 5) for (a) original data, (b) BLS-GSM method, (c) ATW-LM method, (d) MPG-ALS method, (e) SURE-LET method, (f) BPR-RR method (vertical axis: photon counts; horizontal axis: eV).}\label{6}
\end{figure}
\begin{figure}\centering
  \includegraphics[width=96.8mm]{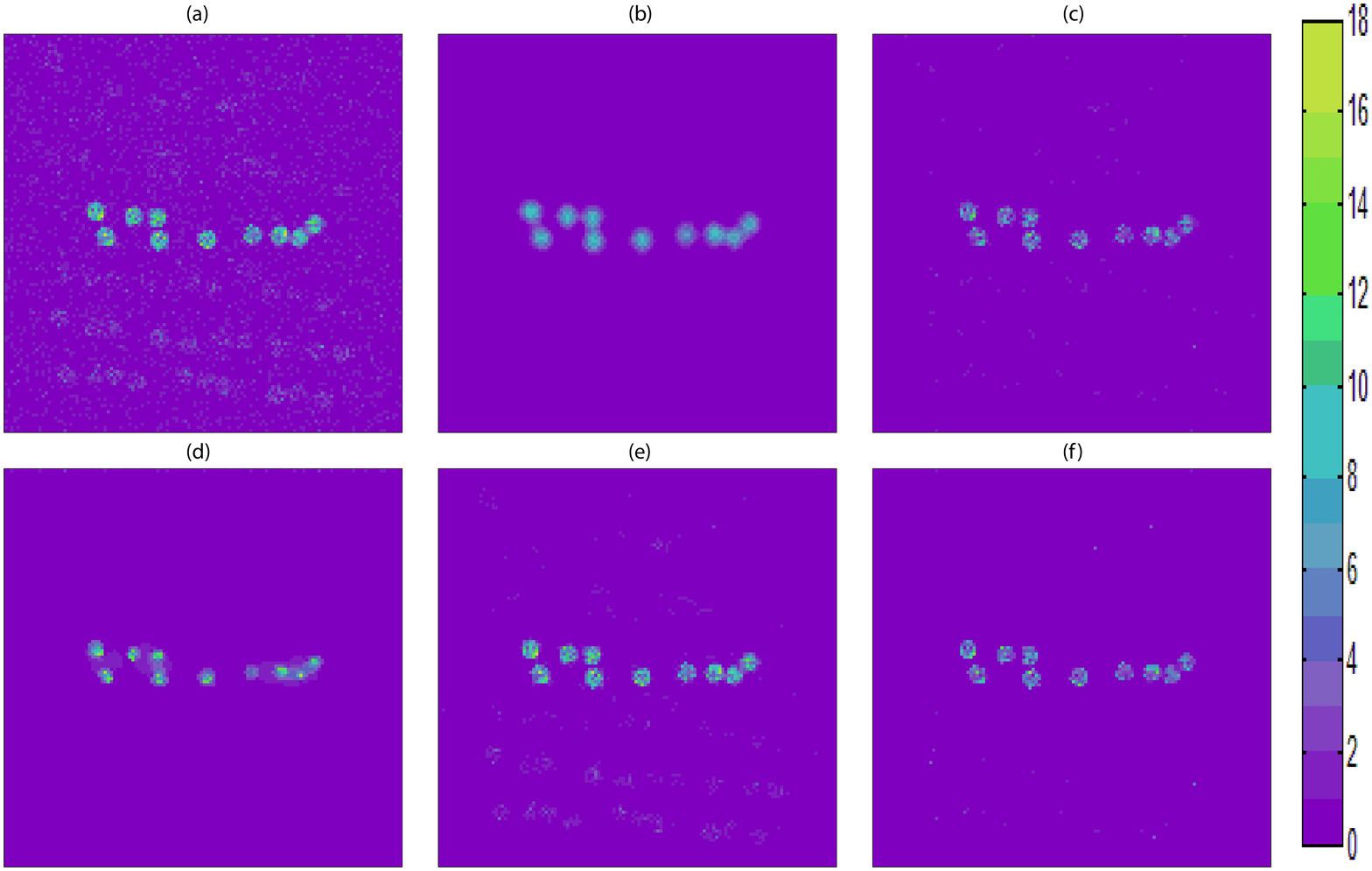}\\
  \caption{The X-ray spectral images at the energy channel of Zn's KL2 Line Energy (8.62 KeV) for (a) original data, (b) BLS-GSM method, (c) ATW-LM method, (d) MPG-ALS method, (e) SURE-LET method, (f) BPR-RR method.}\label{7}
\end{figure}
\begin{figure}\centering
  \includegraphics[width=96.8mm]{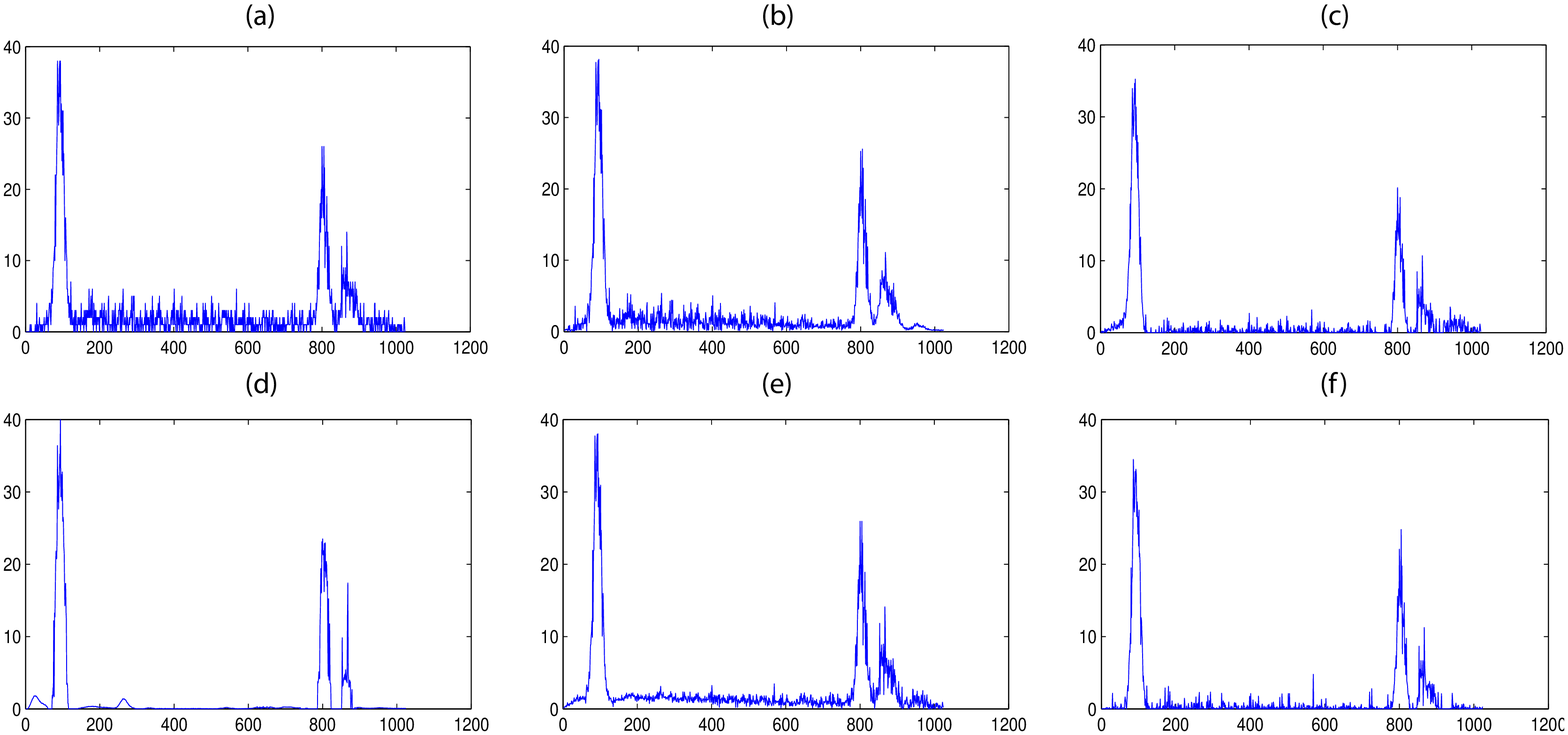}\\
  \caption{Typical single-pixel spectra of Cu/Zn alloy wire (line c in figure 2(a), the line displayed in figure 7) for (a) original data, (b) BLS-GSM method, (c) ATW-LM method, (d) MPG-ALS method, (e) SURE-LET method, (f) BPR-RR method (vertical axis: photon counts; horizontal axis: eV).}\label{8}
\end{figure}
\begin{figure}\centering
  \includegraphics[width=96.8mm]{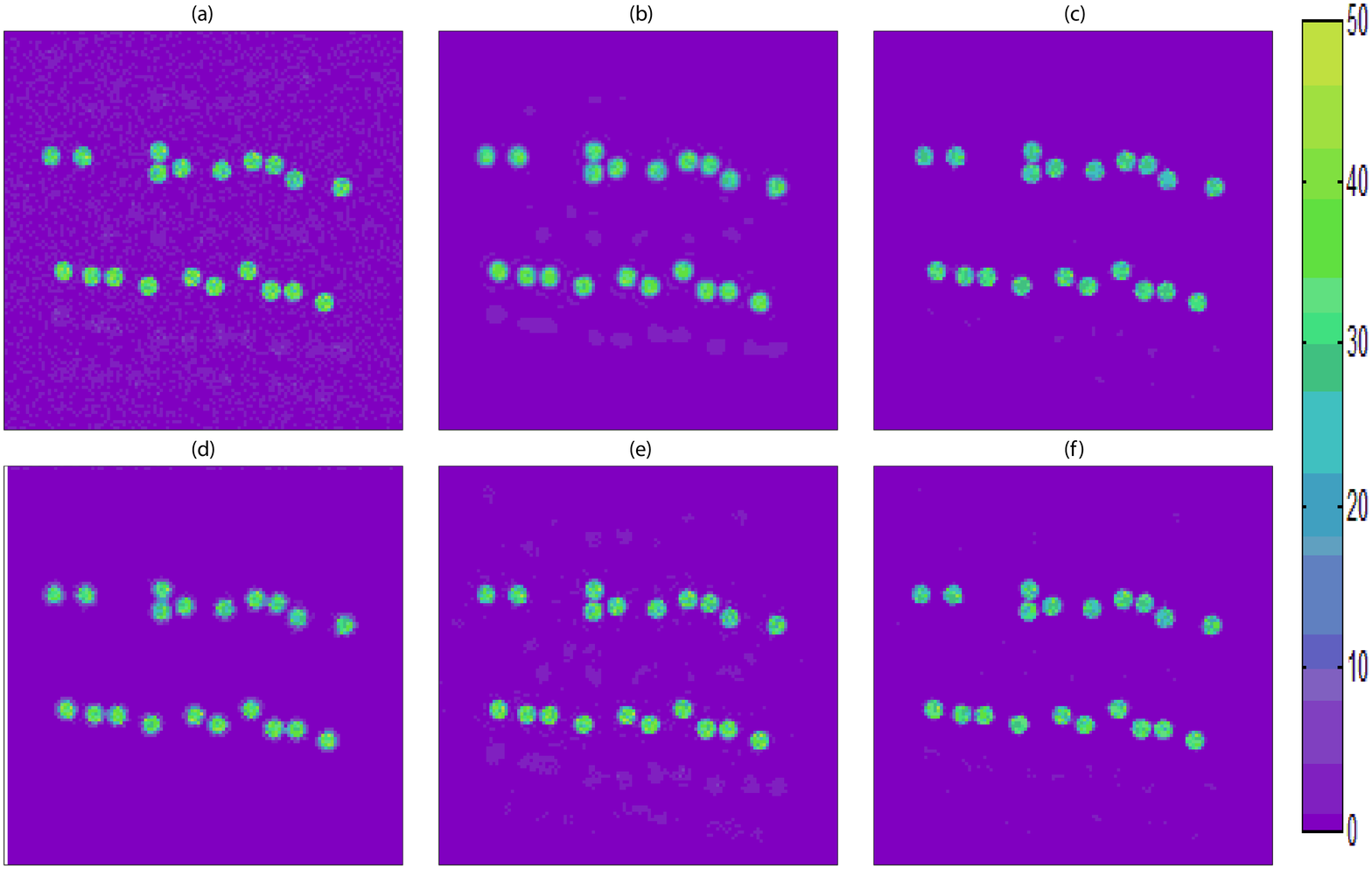}\\
  \caption{The X-ray spectral images at the energy channel of Fe's KL2 Line Energy (6.39 KeV) for (a) original data, (b) BLS-GSM method, (c) ATW-LM method, (d) MPG-ALS method, (e) SURE-LET method, (f) BPR-RR method.}\label{9}
\end{figure}

Since the whole spectra data contain 1024 spectral images, it is not possible to display all of them in this paper. Here we choose the typical spectral images at KL2 line energy channel to show the denoising and baseline drift removal performance with the colour scale representing different numbers of photon counts. We will also give the corresponding 1D plots showing single-pixel spectra for different elements. From the visual inspection aspect, the performance can be evaluated in terms of removing the noises and baseline drift at background regions and preserving the original photon counts at target regions.

The low-concentration elements Mn, Cr and Zn have very weak photon counts and low SNRs in their original data (see figure 3(a), 5(a) and 7(a)), which are more apparent in the corresponding single-pixel spectra (figure 4(a), 6(a) and 8(a)). After denoising, the MPG-ALS method have changed the original photon counts of these low-concentration elements at target regions (see figure 3(d), 5(d) and 7(d)), and the spectral waveforms have distorted (figure 4(d), 6(d), and 8(d)). The BLS-GSM method smoothes the raw spectral data too much as the photon counts at target regions have been homogenized (see figure 3(b), 5(b) and 7(b)) while preserves relatively strong noises in the spectra (figure 4(b), 6(b) and 8(b)). The denoised data by SURE-LET method resemble the original photon count data very much for the positions of elements Mn, Cr and Zn, while the denoising performance is not desirable at background and other component elements' positions (see figure 3(e), 5(e) and 7(e)). SURE-LET method also has the worst baseline drift removing performance compared with the other methods (figure 4(e), 6(e) and 8(e)).

Among all the five methods, BPR-RR and ATW-LM methods achieve the best performance in terms of both denoising and preserving the original photon counts. Figure 3(c)(f), 5(c)(f) and 7(c)(f) show the satisfying denoising effect of BPR-RR and ATW-LM methods for element Mn, Cr and Zn simultaneously with the high fidelity to original photon counts. However, we can see that there are less noise left on the spectra of BPR-RR method than the spectra of ATW-LM method (figure 4(c)(f), 6(c)(f) and 8(c)(f)). Although the BPR-RR method shows a decrease in photon count (intensity) compared to the raw data (figure 7(a) and (f)), this performance is acceptable since the raw data are contaminated with Poisson noise, which means the raw maximum does not always represent the real maximum photon counts. In single-pixel X-ray spectra, the KM peak of Cu (the peak at 890 on figure 8. (a)(f)) and the KL peak of Zn (the peak at 862 on figure 8. (a)(f)) are so close that the peak of Cu will cause an effect of baseline drift on the peak of Zn, which means the energy channel 862 contains two elements (Cu and Zn) rather than one element (Zn). For this reason, BPR-RR method has partly removed the baseline drift effect of Cu and thus causes the loss of denoised photon counts of Zn (figure 7(f)). As for element Fe with high SNR, all the above methods have similar good results except SURE-LET method (see figure 9). The corresponding 1D plots of single-pixel spectra of the lower line in figure 9 is actually displayed in figure 6.

\begin{figure}\centering
  \includegraphics[width=90mm]{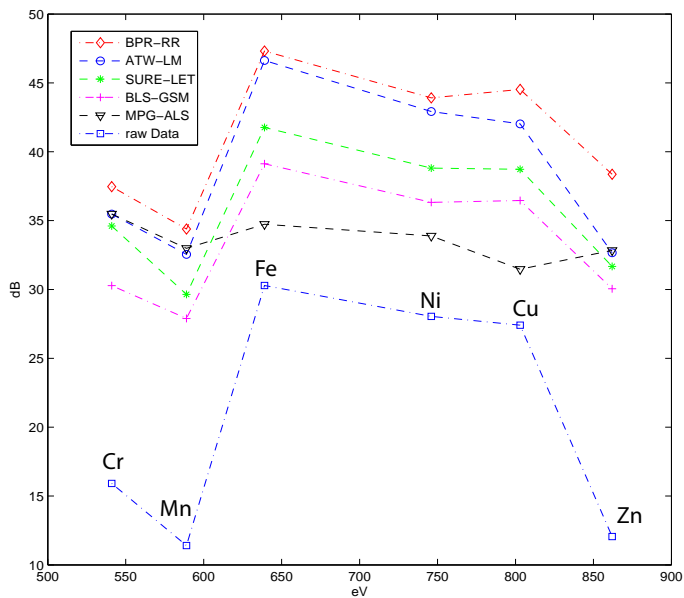}\\
  \caption{The signal-to-noise ratios estimated from the X-ray spectral images of each element's KL2 line energy. The elements from the left to right are Cr (5.41 KeV), Mn (5.89 KeV), Fe (6.39 KeV), Ni (7.46 KeV), Cu (8.03 KeV) and Zn (8.62 KeV).}\label{8}
\end{figure}

Although the noiseless data is unavailable, we decided to estimate the SNR over the whole data. Keenan (2007) has provided a way to calculate the SNR of Poisson data. With this method, the SNR of the original X-ray spectral data matrix $\mathbf{D}$ can be calculated as
\begin{equation}\label{eq21}
    \mathbf{SNR} = \frac{{sum\;of\;eigenvalues\;of\;{\mathbf{D}^T}\mathbf{D}}}{{sum\;of\;all\;data\;elements\;in\;\mathbf{D}}}
\end{equation}
However, the equation (21) can not be used to calculate the SNR of spectra data preprocessed by above-mentioned methods because the preprocessed data have lost their Poisson nature. All of the preprocessed data can be seen as true data corrupted with Gaussian white noise. Since no wire signal is present in the background areas (black areas in figure 2(a)), data of these clear background areas can be taken as noise to calculate the variance of Gaussian noise in the preprocessed data. However, this variance of the clear background area cannot be used as the variance of the ensemble image unless the noise process is ergodic. Actually, the acquisition of whole multidimensional X-ray photon count data and inherent Poisson noise follows a typical Poisson process (Boulanger \etal 2010) with ergodic properties (Wolff 1981). Furthermore, our space-independent denoising algorithm will not change the ergodicity. Therefore, we use the variance of the background area as the noise variance of the ensemble image. By means of standard image processing methods such as threshold, it is easy to pick up the data in the background area. The signal mean is easily calculated by using the data on the areas of wire dots. Finally, the SNR of preprocessed data is computed as a ratio of the signal mean to the noise standard deviation (the square root of noise variance).

Figure 10 shows the SNRs of the images in the KL2 line energy channels of the six component elements. Figure 10 display that component elements with small SNRs still have the relative small SNRs in the processed data. The MPG-ALS method and BPR-RR method can enhance the SNRs of low-concentration elements (Cr, Mn and Zn) to almost the same level as the SNRs of high-concentration elements (Ni, Fe and Cu). The ATW-LM method, SURE-LET method and BPR-RR method have similar high computed SNRs while our method has the highest computed SNRs among these methods. (All the elements' line energy data are referred to the database of the National Institute of Standards and Technology of U. S.)

\subsection{Standard biological sample experiment}
\subsubsection{Experimental data}

The two commercially available standard biological samples, bovine liver (NIST 1577a) and pig liver (GBW 08551) with known element concentrations, are imaged with \si{\micro}SXRF at the beamline BL15U at Shanghai Synchrotron Radiation Facility (Shanghai, China). These two samples were separately pressed into a $\sim$ \SI{5}{\milli\gram\per\square\centi\metre} tablet, and then sandwiched between double Mylar films. The image dimensions of bovine liver and pig liver are $ 11 \times 11$ and $ 10 \times 10$, respectively. Both samples are scanned with a complete 2048-channel spectrum at each pixel. More details about the beamline station and the preparation of samples can be found in Wang \etal (2010).

\subsubsection{Results and analysis}

Based on the results of section 3.1, we choose ATW-LM method, SURE-LET method and our BPR-RR method to preprocess the two standard biological samples. However, SURE-LET method failed to process the spectra data of these samples because the quantity of these two samples data are too small to be processed by SURE-LET method. All the parameters of the other methods were set as in section 3.1.

We further use the method described in Marco \etal (1999) to implement quantitative analysis of these two biological samples. In \si{\micro}SXRF imaging, the varied current intensity of X-ray and the differences in thickness and density of thin biological samples can result in considerable errors and undesirable precision. Thus, for enhancing quantification precision, an internal reference in the test specimen must be used to demonstrate the relationship between the fluorescent intensity of the analyte and a signal from an internal reference. As proposed in Marco \etal (1999), the Compton peak of \si{\micro}SXRF spectra is possible to be used as an internal standard for trace element (e.g., Ca, Fe, Cu and Zn) quantification in organic matter due to the Compton scattering being theoretically related to the mass of the sample. The commercially available standard reference materials, having closely matched matrix with the analyte and known element concentration, are used to compute the sensitivity $R_{ic}$ of element $i$ relative to the Compton peak using the following equation:
\begin{equation}\label{eq22}
    {R_{ic}} = \frac{{{I_{ir}}}}{{{I_{cr}}}}\frac{1}{{{C_{ir}}}}
\end{equation}
where $I_{ir}$ is the fluorescent intensity of each element $i$ in the standard reference sample, $I_{cr}$ is the intensity for the Compton scattering peak of the standard reference sample, $C_{ir}$ is the known element concentrations of the standard reference sample.

With the relative ratio $R_{ic}$ of the sensitivity of each element $i$, the concentration $C_{ia}$ of each element in the analyte can be calculated by the intensity of each element $i$ and Compton scattering peak area of the analyte, {\it i.e.}
\begin{equation}\label{eq23}
    {C_{ia} = \frac{{{I_{ia}}}}{{{I_{ca}}}}\frac{1}{{{R_{ic}}}}}
\end{equation}
where $I_{ia}$ is the fluorescent intensity of each element $i$ in the analyte, $I_{ca}$ is the intensity for the Compton scattering peak of the analyte.

Based on the above-described method, we choose the bovine liver as the standard reference sample with the matrix-matched pig liver being treated as the standard analyte to valuate our proposed method. According to the element compositions in these two standard biological samples, we choose the elements Ca, Fe, Cu and Zn to perform the quantitative calculations. Firstly, we use three different bovine liver data, {\it i.e.}, the BPR-RR, ATW-LM method preprocessed bovine liver data and the raw bovine liver data, to calculate the three different sets of the intensities for each element $I_{ir}$ and Compton peak $I_{cr}$ by peak area integration (figure 1(c)), and then compute the three different sets of sensitivities $R_{ic}$ of element $i$ relative to the Compton peak according to the equation (22) (see table 1). Although the same bovine liver sample making specimen matrices and element contents constant in the three preprocessed data, the different preprocessing methods with (or without) different Poisson denoising and baseline drift removal methods can affect the relative sensitivity $R_{ic}$.

Secondly, we use the BPR-RR, ATW-LM method preprocessed pig liver data and the raw pig liver data to calculate the three different sets of intensities for each element $I_{ia}$ and the Compton peak $I_{ca}$. Then by using the $R_{ic}$ calculated from the matrix-matched bovine liver data, the concentrations of Ca, Fe, Cu and Zn in the standard pig liver data were calculated according to the equation (23) and in comparison with the certified values in table 2.

All the intensities for the selected elements and Compton scattering peak are calculated by PyMCA. Although PyMCA itself has preprocessing methods, here we only use this software to separate and integrate different elements' characteristic peaks and Compton peaks in the raw liver data, BPR-RR and ATW-LM method preprocessed liver data.

\begin{table}
\caption{\label{tabone}The ratios ${R_{ic}}$ obtained from the raw bovine liver (NIST 1577a) data and the data preprocessed by the ATW-LM and BPR-RR methods}

\begin{indented}
\lineup
\item[]\begin{tabular}{@{}*{4}{l}}
\br
&\centre{3}{${R_{ic}} \times {10^2}$}\\
\crule{4}\\
\0\0    &  raw data  &  ATW-LM  &  BPR-RR  \cr
\mr
\0\0Ca&\centre{1}{$0.75\pm0.04$}&\centre{1}{$0.69\pm0.04$}&\centre{1}{$0.77\pm0.05$}\cr
\0\0Fe&\centre{1}{$1.47\pm0.15$}&\centre{1}{$1.50\pm0.15$}&\centre{1}{$1.59\pm0.16$}\cr
\0\0Cu&\centre{1}{$2.73\pm0.12$}&\centre{1}{$2.68\pm0.12$}&\centre{1}{$2.86\pm0.13$}\cr
\0\0Zn&\centre{1}{$3.36\pm0.22$}&\centre{1}{$3.47\pm0.23$}&\centre{1}{$3.62\pm0.24$}\cr
\br
\end{tabular}
\end{indented}
\end{table}

\begin{table}
\caption{\label{tabone}The concentrations of elements (\si{\micro\gram\per\gram}) in pig liver data (GBW 08551) calculated by using the different ${R_{ic}}$ from the raw and the two preprocessed bovine liver data (NIST 1577a) at table 1. }

\begin{indented}
\lineup
\item[]\begin{tabular}{@{}*{5}{l}}
\br
&\centre{4}{Concentrations of elements: \si{\micro\gram\per\gram}}\\
\crule{5}\\
\0\0  &Certified value&raw data&ATW-LM&BPR-RR\cr
\mr
\0\0Ca&\centre{1}{$197\pm14$}&\centre{1}{$256\pm15$}&\centre{1}{$235\pm14$}&\centre{1}{$195\pm11$}\cr
\0\0Fe&\centre{1}{$1050\pm40$}&\centre{1}{$1144\pm118$}&\centre{1}{$1062\pm97$}&\centre{1}{$1055\pm109$}\cr
\0\0Cu&\centre{1}{$17.2\pm1.0$}&\centre{1}{$16.2\pm0.7$}&\centre{1}{$16.5\pm0.7$}&\centre{1}{$17.4\pm0.8$}\cr
\0\0Zn&\centre{1}{$172\pm8$}&\centre{1}{$199\pm13$}&\centre{1}{$167\pm8$}&\centre{1}{$178\pm12$}\cr
\br
\end{tabular}
\end{indented}
\end{table}

Table 2 shows that Poisson denoising and baseline drift removal methods do have positive effect on the accuracy and precision for quantitative analysis of the different trace elements. Among the three different preprocessing methods, BPR-RR method has produced best quantitative analysis of element concentrations that are in good agreement with the certified values. The BPR-RR method has obtained high accuracy and high precision for quantitative analysis of the trace element Ca and Cu. Although the quantitative analysis of Zn is not as good as those of Ca and Cu, the BPR-RR method still has good accuracy and good precision for quantitative analysis of trace element Zn. The possible cause of the decrease of accuracy and precision for quantitative analysis of Zn is already mentioned in the section 3.1.2. Both bovine liver and pig liver contain the elements of Cu and Zn. The peaks of the spectra of Cu and Zn are so close that even an excellent tool such as PyMCA is not able to separate these two elements in the spectra thoroughly and precisely. As for the element Fe, the BPR-RR method has achieved good accuracy but poor precision for quantitative analysis due to the calculated standard deviation uncertainty ($\pm109$) being significantly larger than the certified standard deviation uncertainty ($\pm40$). One explanation for this is that the concentration of element Fe in pig liver (GBW 08551) is very high to introduce large photon counts in the spectrum. Large photon counts assume an approximately normal distribution rather than Poisson distribution (Haight 1967). Therefore, the method that is designed to deal with the low photon count data ({\it e.g.} Cu, Ca and Zn) will not have the excellent denoising effect for the high photon count data of element Fe at pig liver.

\section{Conclusion}
In this paper, we have introduced a bootstrap Poisson regression and robust nonparametric regression method to reduce Poisson noise and baseline drift in the multidimensional X-ray spectral data. The proposed method is very effective to improve the SNRs of the raw X-ray spectral data. The comparison with other competing methods shows that the BPR-RR method offers performance better than some state-of-the-art approaches. By using two standard biological samples and applying Compton peak standardization in \si{\micro}SXRF quantitative imaging, the BRP-RR preprocessing method can ensure satisfying accuracy and precision for quantitative analysis of the trace elements in biological samples. The concentrations of Ca, Fe, Cu and Zn in the standard reference material (GBW 08551, pig liver) determined by BRP-RR preprocessing and subsequent quantitative imaging method were in good agreement with the certified values. This work can be extended along several directions in the future. First, the bootstrap samples can be sampled through nonparametric ways to converge more quickly to the statistic features of the original raw data. Second, Poisson regression can be computed within the Bayesian framework, where a priori information about the expectation of photon count data can be introduced to improve its performance. Third, our algorithm cannot achieve satisfactory performance in the case of extreme local discontinuities (or local nonhomogeneity) at all directions from the current scanning point of interest in the sample, though, such case is a rare occurrence in modern high-resolution biomedical X-ray spectral imaging. At last, the Poisson denoising and baseline removal method are automatically adaptive not only to the low-concentration elements but also to the high-concentration elements that produce high photon count data (such as the element Fe in liver data), while our proposed method is apt to deal with the low photon count data.

\ack
The author would like to thank Dr. Paul Kotula for providing the wires' EDS dataset. This work was partially performed at the Shanghai Synchrotron Radiation Facility (SSRF) in China and supported by the National Basic Research Program of China (973 Program 2010CB834300, 2011CB933403), National Natural Science Foundation of China (61271320, 21175136 and 60872102), China Scholarship Council and the small animal imaging project (06-545). The authors would like to thank the anonymous reviewers who contributed to considerably improve the quality of this paper.

\section*{Reference}
\begin{harvard}
    \item[]Anscombe F J 1948 The transformation of Poisson, binomial and negative-binomial data {\it Biometrika} {\bf 35} 246-54
    \item[]Bekemans B, Vincze L, Somogyi A, Drakopoulos M, Kempenaers L, Simionovici A and Adams F 2003 Quantitative X-ray fluorescence analysis at the ESRF ID18F microprobe {\it Nuclear Instruments and Methods in Physics Research B} {\bf 199} 396-401
    \item[]Bonettini S and Ruggiero V 2011 An alternating extragradient method for total variation-based image restoration from Poisson data {\it Inverse Problems} {\bf 27} 095001
    \item[]Boulanger J, Kervrann C, Bouthemy P, Elbau P, Sibarita J B and Salamero J 2010 Patch-Based Nonlocal Functional for Denoising Fluorescence Microscopy Image Sequences {\it IEEE Trans. Medical Imaging} {\bf 29} 442-54
    \item[]Boulanger J, Sibarita J B, Kervrann C and Bouthemy P 2008 Non-parametric regression for patch-based fluorescence microscopy image sequence denoising {\it 5th IEEE Int. Sym. Biomedical Imaging} 748-51
    \item[]Chan R H and Chen K 2007 Multilevel algorithm for a Poisson noise removal model with total-variation regularization {\it Int J. Comput. Math.} {\bf 84} 1183-198
    \item[]Cleveland W S 1979 Robust locally weighted regression and smoothing scatterplots {\it Journal of the American Statistical Association} {\bf 74} 829-36
    \item[]Dahlbom M 2002 Estimation of image noise in PET using the bootstrap method {\it IEEE Trans. Nuclear Science} {\bf 49} 2062-66
    \item[]Davison A C and Hinkley D V 1997 {\it Bootstrap Methods and their Application} (Cambridge University Press) chap.7 p~327
    \item[]Efron B 1979 Bootstrap methods: another look at the jackknife {\it the annals of statistics} {\bf 7} 1-26
    \item[]Eilers P H 2004 Parametric Time Warping {\it Anal. Chem.} {\bf 76} 404-11
    \item[]Friedrichs M S 1995 A model-free algorithm for the removal of baseline artifacts {\it Journal of Biomolecular NMR} {\bf 5} 147-53
    \item[]Geraki K, Farquharson M J and Bradley D A 2004 X-ray fluorescence and energy dispersive x-ray diffraction for the quantification of elemental concentrations in breast tissue {\it Phys. Med. Biol.} {\bf 49} 99-110
    \item[]Gherase M R and Fleming D E 2011 A calibration method for proposed XRF measurements of arsenic and selenium in nail clippings {\it Phys. Med. Biol.} {\bf 56} 215-25
    \item[]Haight F A 1967 {\it Handbook of the Poisson Distribution} (New York: John Wiley \verb"&" Sons)
    \item[]Haynor D R and Woods S D 1989 Resampling estimates of precision in emission tomography {\it IEEE Trans. Medical Imaging} {\bf 8} 337-43
    \item[]Jenkins R, Gould R W and Gedcke D 1995 Quantitative X-Ray Spectrometry 2nd ed {\it Marcel Dekker; New York}
    \item[]Keenan M R 2007 Multivariate Analysis of Spectral Images Composed of Count Data {\it Techniques and Applications of Hyperspectral Image Analysis} (West Sussex, England: John Wiley \& Sons Ltd) chapter 5
    \item[]Kervrann C and Trubuil A 2004 An adaptive window approach for Poisson noise reduction and structure preserving in confocal microscopy {\it IEEE Int. Sym. Biomedical Imaging} {\bf 1} 788-91
    \item[]Kolaczyk E D 2000 Nonparametric estimation of intensity maps using Haar wavelets and Poisson noise characteristics {\it Astrophys. J.} {\bf 534} 490-505
    \item[]Kolaczyk E D 1999 Bayesian multiscale models for Poisson processes {\it J. Amer. Statist. Assoc.} {\bf 94} 920-33
    \item[]Kotula P G, Keenan M R and Michael J R 2003 Automated analysis of SEM X-ray spectral images: a powerful new microanalysis tool {\it Microscopy and Microanalysis} {\bf 9} 1-17
    \item[]Le T, Chartrand R, and Asaki T J 2007 A variational approach to reconstructing images corrupted by Poisson noise {\it J. Math. Imaging Vis.} {\bf 27} 257-263
    \item[]Lefkimmiatis S, Maragos P and Papandreou G 2009 Bayesian inference on multiscale models for Poisson intensity estimation: applications to photon-limited image denoising {\it IEEE Trans. Image Proc.} {\bf 18} 1724-41
    \item[]Liland K H, Almoy T and Mevik B H 2010 Optimal Choice of Baseline Correction for Multivariate Calibration of Spectra {\it Applied Spectroscopy} {\bf 64} 1007-16
    \item[]Luisier F and Blu T 2008 SURE-LET Multichannel Image Denoising: Interscale Orthonormal Wavelet Thresholding {\it IEEE Trans. Image Processing} {\bf 17} 482-92
    \item[]Makitalo M and Foi A 2011 Optimal Inversion of the Anscombe Transformation in Low-Count Poisson Image Denosing {\it IEEE Trans. Image Processing} {\bf 20} 99-109
    \item[]Marco L M, Greaves E D and Alvarado J 1999 Analysis of human blood serum and human brain samples by total reflection X-ray fluorescence spectrometry applying Compton peak standardization {\it Spectrochimica Acta Part B: Atomic Spectroscopy} {\bf 54} 1469-1480
    \item[]Mounicou S, Szpunar J and Lobinski R 2009 Metallomics: the concept and methodology {\it Chemical Society Reviews} {\bf 38} 1119-38
    \item[]Palakkal S and Prabhu K M M, 2012 Poisson image denoising using fast discrete curvelet transform and wave atom {\it Signal Processing} {\bf 92} 2002-17
    \item[]Popescu B F, George M J, Bergmann U, Garachtchenko A V, Kelly M E, McCrea R P, Luning K, Devon R M, Geoge G N, Hanson A D, Harder S M, Chapman L D, Pickering I J and Nichol H 2009 Mapping metals in Parkinson's and normal brain using rapid-scanning x-ray fluorescence {\it Phys. Med. Biol.} {\bf 54} 651-63
    \item[]Qin Z, Toursarkissian B and Lai Barry 2011 Synchrotron radiation X-ray fluorescence microscopy reveals a spatial association of copper on elastic laminae in rat aortic media {\it Metallomics} {\bf 3} 823-28
    \item[]Ruckstuhl A F, Jacobson M P, Field S W and Dodd J A 2001 Baseline subtraction using robust local regression estimation {\it Journal of Quantitative Spectroscopy and Radiative Transfer} {\bf 68} 179-93
    \item[]Sole V A, Papillon E, Cotte M, Walter Ph and Susini J 2007 A multiplatform code for the analysis of energy-dispersive X-ray fluorescence spectra {\it Spectrochimica Acta Part B: Atomic Spectroscopy} {\bf 62} 63-68
    \item[]Spring B Q and Clegg R M 2009 Image analysis for denoising full-field frequency-domain fluorescence lifetime images {\it Journal of Microscopy} {\bf 235} 221-37
    \item[]Timmerman K E and Nowak R D 1999 Multiscale modeling and estimation of Poisson processes with application to photon-limited imaging {\it IEEE Trans. Inf. Theory} {\bf 45} 846-62
    \item[]Twining B S, Baines S B, Fisher N S, Maser J, Vogt S, Jacobsen C, Tovar-Sanchez A and Sanudo-Wilhelmy S A 2003 Quantifying Trace Elements in Individual Aquatic Protist Cells with a Synchrotron X-ray Fluorescence Microprobe {\it Anal. Chem.} {\bf 75} 3806-16
    \item[]Van Grieken R E and Markowicz A A 2002 {\it Handbook of X-ray Spectrometry 2nd} (New York: Marcel Dekker) chapter 3
    \item[]Wang H J, Wang M, Wang B, Meng X Y, Wang Y, Li M, Feng W Y, Zhao Y L and Chai Z F 2010 Quantitative imaging of element spatial distribution in the brain section of a mouse model of Alzheimer's disease using synchrotron radiation X-ray fluorescence analysis {\it J. Anal. At. Spectrom.} {\bf 25} 328-33
    \item[]Wang Y L 2007 Noise-induced systematic errors in ratio imaging: serious artifacts and correction with multi-resolution denoising {\it Journal of Microscopy} {\bf 228} 123-31
    \item[]Wolff R W 1981 Poisson Arrivals See Time Averages {\it Operations Research} {\bf 30} 223-31
    \item[]Zhang B, Fadili J M and Starck J L 2008 Wavelets, Ridgelets and Curvelets for Poisson Noise Removal {\it IEEE Transactions on Image Processing} {\bf 17} 1093-108
    \item[]Zhou W and Li Q 2013 Adaptive total variation regularization based scheme for Poisson noise removal {\it Math. Meth. Appl. Sci.} {\bf 36} 290-99
\end{harvard}

\end{document}